# Accelerating key bioinformatics tasks 100-fold by improving memory access

Igor Sfiligoi
University of California San Diego
La Jolla, CA, USA
isfiligoi@sdsc.edu

Daniel McDonald
University of California San Diego
La Jolla, CA, USA
danielmcdonald@ucsd.edu

Rob Knight
University of California San Diego
La Jolla, CA, USA
robknight@ucsd.edu

## ABSTRACT

Most experimental sciences now rely on computing, and biological sciences are no exception. As datasets get bigger, so do the computing costs, making proper optimization of the codes used by scientists increasingly important. Many of the codes developed in recent years are based on the Python-based NumPy, due to its ease of use and good performance characteristics. The composable nature of NumPy, however, does not generally play well with the multi-tier nature of modern CPUs, making any non-trivial multi-step algorithm limited by the external memory access speeds, which are hundreds of times slower than the CPU's compute capabilities. In order to fully utilize the CPU compute capabilities, one must keep the working memory footprint small enough to fit in the CPU caches, which requires splitting the problem into smaller portions and fusing together as many steps as possible. In this paper, we present changes based on these principles to two important functions in the scikit-bio library, principal coordinates analysis and the Mantel test, that resulted in over 100x speed improvement in these widely used, general-purpose tools.

## CCS CONCEPTS

• Software and its engineering~Software notations and tools~Software libraries and repositories • Applied computing~Life and medical sciences~Bioinformatics

## 1 Introduction

The Python language [1] has been adopted by many sciences for their computational needs, due to its ease of use and readability. While Python native data structures tend to be relatively slow, several standard libraries that provide high performance vector and matrix operations have been developed, with NumPy [2,14] being one popular choice. NumPy is both easy to use and very fast for the operations it implements. Many science problems, however, need more that a single NumPy operation to get scientific insights, resulting in several independent incremental steps. As datasets become bigger, intermediate memory representations can easily exceed the cache size of the CPU in use, making an algorithm memory bound, and thus sub-optimal.

One Python-based scientific library in bioinformatics is scikit-bio [3], which is a core library for popular microbiome analysis packages such as QIIME [4]. Two of heavily used, yet computationally expensive functions in that library are `pcoa`, which performs a principal coordinate analysis [10] of a distance matrix, and `mantel`, which computes the correlation between distance matrices using the Mantel test [10]. These functions became critical bottlenecks following the upstream optimization of UniFrac distance calculations [5]. A careful examination of the NumPy-based source code revealed the memory-bound nature of the implementation, providing insight in the design of new implementations that are between 30x and 200x faster than the original on any non-trivial input.

The structure of this paper is as follows. Section 2 provides an overview of modern CPU architectures, with an emphasis on the multi-tiered memory system. Section 3 provides an overview of the microbiome science, with an emphasis of how the above-mentioned functions are being used in the analysis pipeline. And Section 4 provides the analysis of the `pcoa` and `mantel` implementation before and after the memory-aware optimization.

### 1.1 Related work

The importance of cache-optimized algorithms has been known for a very long time [6-7] and is the basis of most high-performance libraries maintained by computing professionals in lower-level languages.

In this paper we focus on code that is maintained by science users, which tends to prioritize ease of programming and readability, and is often written in higher-level languages, like

Python with NumPy. While there are legitimate reasons to go that route, we aim to show both the drawbacks of this approach and provide a possible path forward.

## 2 The multi-tier memory system

The importance of cache-optimized algorithms has been known for a very long time [6-7] and is the basis of most high-performance libraries maintained by computing professionals in lower-level languages.

Due to both cost and manufacturing complexity, cache memories must balance latency with capacity. In the tested 16 core Intel Xeon CPU, the fastest L1 cache provides a 4 cycle latency, but has the capacity of only 32 KiB per core [9]. The most useful L2 cache is 1 MiB per core, but already incurs a latency penalty of 14 cycles. There is also a shared 22.5 MiB L3 cache, with a latency of 50+ cycles. Other CPUs, e.g. the AMD EPYC series, have slightly different size and latency characteristics, but are still in the same ballpark. Cache size wise, the tested 8 core AMD EPYC 7252 CPU has 96 KiB of L1 cache per core, 512 KiB L2 cache per core and a shared 64 MiB L3 cache.

Any algorithm that wants to minimize the time to solution must thus keep access to DRAM as low as possible. This implies that any multi-step application must utilize buffers that fit in the CPU caches, ideally in the fastest but small L1 cache.

One further complication is the fact that caches replace their content not in single bytes, but rather much longer contiguous buffers, known as cache lines; in the case of the Intel Xeon CPU, the cache line is 64 bytes, i.e. 512 bits, across all cache levels. Algorithms accessing non-contiguous memory areas must thus take that into account when estimating memory use.

## 3 The microbiome science compute challenge

A microbiome study may be composed of a large number of samples from the real world. In the Earth Microbiome Project (EMP) [15], for example, tens of thousands of samples were collected from many different environments (e.g., marine sediment, animal feces, plant surfaces, glaciers, hot springs, etc) by over 100 research labs from all over the world. Microbiome studies, including the EMP, typically utilize DNA sequencing technology in order to observe what microorganisms are present in a given sample. From these observations, one of first questions researchers ask is how similar the samples are to each other, and can these similarities be explained by study variables (e.g., salinity, pH, host or not host associated, etc).

The computation of sample similarity is performed pairwise producing a distance matrix of size N x N where N is the number of samples being evaluated. Following construction of a distance matrix, researchers often apply principal coordinates analysis, a dimensionality reduction technique similar to principal components analysis, for visualization and statistical purposes. In addition, researchers also frequently wish to ask whether one distance matrix is correlated to another through the use of a non-parametric Mantel test; there are many ways to construct a distance matrix, and many ways to assess the composition of a microbiome sample.

Recently, the authors optimized a common distance metric, UniFrac [5], to scale to datasets with hundreds of thousands of samples. However, the types of operations that could be successfully applied to these distance matrices were limited, motivating investigation into the optimization of downstream methods.

## 4 Optimizing scikit-bio distance matrix operations

The scikit-bio library is widely used in the bioinformatics community and contains many useful classes and functions. It is written mostly in Python and relies heavily on NumPy for the most compute intensive parts.

For the purpose of this document, we will focus on only two specific functions operating on distance matrix objects, namely `pcoa` and `mantel`. Each will be described separately, with an analysis of the implementation before and after the memory-focused optimization.

### 4.1 Principal Coordinates Analysis

The `pcoa` function is used to perform a Principal Coordinates Analysis of a distance matrix. The function is composed roughly of two parts: centering of the matrix [10], and compute of approximate eigenvalues [11]. Unexpectedly, the original implementation spent most of the time in the first step, i.e. centering the matrix.

The used implementation of the centering the matrix is summarized in Algorithm 1. The implemented algorithm is simple and elegant but makes no attempt at being memory efficient. Each arithmetic operation traverses the whole NumPy matrix, which for most problems is larger than any CPU cache. Moreover, it traverses the matrix multiple times to get the various needed mean values, which will again be memory bound. In total, the algorithm will read 8 and write 5 matrix-sized buffers into the off-chip memory, alongside 2 much smaller one-dimensional arrays containing the mean values.

---

Algorithm 1: Original center_distance_matrix implementation

```
import numpy as np

def e_matrix(distance_matrix):
    return distance_matrix * distance_matrix / -2

def f_matrix(E_matrix):
    row_means = E_matrix.mean(axis=1, keepdims=True);
    col_means = E_matrix.mean(axis=0, keepdims=True)
    matrix_mean = E_matrix.mean()
    return E_matrix - row_means - col_means + matrix_mean

def center_distance_matrix(distance_matrix):
    # distance_matrix must be of type np.ndarray and
    # be a 2D floating point matrix
    return f_matrix(e_matrix(distance_matrix))
```

---

Unfortunately, one cannot do much better by simply concatenating NumPy operations. We thus re-implemented the algorithm using Cython [12], which allows for efficient inner-loop concatenation of arithmetic operations on NumPy matrix data. We also take advantage of the fact that the matrix is symmetric and compute the mean values only in one dimension.

The cache-optimized implementation is presented in Algorithm 2. The code is longer and a little less easy to read, but it only reads and writes 2 times the matrix-sized buffer from off-chip memory, alongside the single one-dimensional array.

---

Algorithm 2: Cache-optimized center_distance_matrix implementation

---

```
import numpy as np
from cython.parallel import prange

def e_matrix_means_cy(float[:, :] mat,
                      float[:, :] centered, float[:] row_means):
    n_samples = mat.shape[0]
    global_sum = 0.0
    for row in prange(n_samples, nogil=True):
        row_sum = 0.0
        for col in range(n_samples):
            val = mat[row,col];
            val2 = -0.5*val*val;
            centered[row,col] = val2;
            row_sum = row_sum + val2

        global_sum += row_sum
        row_means[row] = row_sum/n_samples

    return (global_sum/n_samples)/n_samples

def f_matrix_inplace_cy(float[:] row_means,
                        float global_mean, float[:, :] centered):
    n_samples = centered.shape[0]
    # use a tiled pattern to maximize locality of row_means,
    #since they are also column means
    for trow in prange(0, n_samples, 16, nogil=True):
        trow_max = min(trow+16, n_samples)
        for tcol in range(0, n_samples, 16):
            tcol_max = min(tcol+16, n_samples)
            for row in range(trow, trow_max, 1):
                gr_mean = global_mean - row_means[row]
                for col in range(tcol, tcol_max, 1):
                    centered[row,col] = centered[row,col] +
                        (gr_mean - row_means[col])

def center_distance_matrix(mat, centered):
    # mat and centered must be of type np.ndarray
    # be 2D floating point matrices
    row_means = np.empty(mat.shape[0])
    global_mean = e_matrix_means_cy(mat, centered, row_means)
    f_matrix_inplace_cy(row_means, global_mean, centered)
```

With the reduced off-chip memory traffic, the new algorithm is significantly faster than the original. As can be seen from Table 1, the speedup ranges from about 10x on modest matrix sizes to 30x on really large matrix sizes. Note that the original, NumPy-based version barely benefitted from using multiple cores.

**Table 1: Runtimes for center_distance_matrix, in seconds**

| CPU, number of used cores, matrix size | Original | Latest |
|---|---|---|
| AMD EPYC 7252 1 core - 25k x 25k | 4.1 | 2.1 |
| AMD EPYC 7252 8 cores - 25k x 25k | 4.0 | 0.3 |
| Intel Xeon Gold 6242 1 core - 25k x 25k | 7.7 | 2.3 |
| Intel Xeon Gold 6242 16 cores - 25k x 25k | 7.2 | 0.3 |
| AMD EPYC 7252 1 core - 70k x 70k | 100 | 30 |
| AMD EPYC 7252 8 cores - 70k x 70k | 125 | 4.2 |
| Intel Xeon Gold 6242 1 core - 100k x 100k | 255 | 78 |
| Intel Xeon Gold 6242 16 cores - 100k x 100k | 254 | 9.1 |

## 4.2 Mantel test

The `mantel` function is used to compute the correlation between two distance matrices using the Mantel test. As such, it takes two matrices as input parameters. The statistical significance of a Mantel test is derived from a Monte-Carlo procedure, where the input matrices are permuted K times, followed by a correlation assessment of the elements in each permutation to obtain a null distribution. The most used correlation function is the Pearson correlation coefficient [13], so we will concentrate on it in this paper. The default value for K is 999.

The original implementation of the algorithm was straightforward and is summarized in Algorithm 3. Because the correlation function, namely SciPy's `pearsonr` [16], is used as a black box, the input matrices must be read from main memory at least 1000 times each, along the matrix-sized intermediate buffers, which made this algorithm extremely slow.

---

Algorithm 3: Original mantel implementation

---

```
import numpy as np
from scipy.stats import pearsonr

def _mantel_stats_pearson(x, y, permutations=999):
    # pearsonr operates on the flat representation
    # of the upper triangle of the symmetric matrix
    x_flat = x.condensed_form();
    y_flat = y.condensed_form()
    orig_stat = pearsonr(x_flat, y_flat)

    perm_gen = ( pearsonr(x.permute(condensed=True), y_flat)
                 for _ in range(permutations) )
    permuted_stats = np.fromiter(perm_gen, np.float,
                                 count=permutations)

    return [orig_stat, permuted_stats]

def mantel(x, y, permutations=999):
    orig_stat, permuted_stats = _mantel_stats_pearson(x, y,
                                                       permutations)
    count_better = ( np.absolute(permuted_stats)
                     >= np.absolute(orig_stat) ).sum()
    p_value = (count_better + 1) / (permutations + 1)

    return orig_stat, p_value
```

As before, the key to speeding up memory-bound algorithms is to move as much computation as possible in the inner loop. In order to do that, one has to look into the implementation of the correlation function itself and re-implement it in a multi-matrix fashion, alongside the inlining of the permutation logic.

The implementation of `pearsonr` from the SciPy library is summarized in Algorithm 4, and is very simple.

---

Algorithm 4: Scipy implementation of the Pearson correlation coefficient

---

```
import numpy as np
from linalg import norm

def pearsonr(x_flat, y_flat):
    xm = x_flat – x_flat.mean();
    ym = y_flat – y_flat.mean()
    normxm = norm(xm);
    normym = norm(ym);
    xnorm = xm/normxm;
    ynorm = ym/normym

    return np.dot(xnorm, ynorm)
```

Knowing how `personr` is being invoked, two optimizations that cut the compute needs in half become apparent:

1. Since the 2nd parameter, aka `y_flat`, is always the same, one can pre-compute the transformation once, and just pass the result, aka `ynorm`, to the function.
2. The mean and norm of a matrix do not change when the matrix is permuted, so we can pre-compute those on the original matrix, and re-use them for all successive calls.

With the above optimizations in mind, the task becomes simply the implementation of a permuted multi-matrix dot operation, keeping memory locality in mind. We will not provide the detailed implementation of the original permutation function but do point out that it uses the NumPy provided random permutation function to select the permutation indexes, which we use in the new implementation, too, as summarized in Algorithm 5. As before, Cython is used to implement the expanded loop.

**Table 2: Runtimes for mantel, in seconds**

| CPU, number of used cores, matrix size | Original | Latest |
|---|---|---|
| AMD EPYC 7252 1 core - 10k x 10k | 2.19k | 149 |
| AMD EPYC 7252 8 cores - 10k x 10k | 2.17k | 24 |
| Intel Xeon Gold 6242 1 core - 10k x 10k | 4.2k | 170 |
| Intel Xeon Gold 6242 16 cores - 10k x 10k | 4.2k | 20 |
| AMD EPYC 7252 1 core - 20k x 20k | 8.9k | 615 |
| AMD EPYC 7252 8 cores - 20k x 20k | 8.8k | 97 |
| Intel Xeon Gold 6242 1 core - 20k x 20k | 17.9k | 933 |
| Intel Xeon Gold 6242 16 cores - 20k x 20k | 17.5k | 108 |
| AMD EPYC 7252 8 cores - 70k x 70k | - | 1.44k |
| Intel Xeon Gold 6242 16 cores - 100k x 100k | - | 4.2k |

---

Algorithm 5: Cache-optimized mantel implementation

---

```
import numpy as np
from linalg import norm

def mantel_perm_cy(float[:, :] x_data, float[:, :] perm_order,
                   float xmean, float normxm,
                   float[:] ym_normalized,
                   float[:] permuted_stats):
    perms_n = perm_order.shape[0];
    out_n = perm_order.shape[1]

    for p in prange(perms_n, nogil=True):
        my_ps = 0.0
        for row in range(out_n-1):
            vrow = perm_order[p, row]
            idx = row*(out_n-1) - ((row-1)*row)/2
            for icol in range(out_n-row-1):
                col = icol+row+1
                yval = ym_normalized[idx+icol]
                xval = x_data[vrow, perm_order[p, col]]*mul + add
                my_ps = yval*xval + my_ps

        permuted_stats[p] = my_ps

def _mantel(x, y, permutations):
    x_flat = x.condensed_form();
    y_flat = y.condensed_form()
    xm = x_flat – x_flat.mean();
    ym = y_flat – y_flat.mean()
    normxm = norm(xm);
    normym = norm(ym);
    xnorm = xm/normxm;
    ynorm = ym/normym
    orig_stat = np.dot(xnorm, ynorm)
    mat_n = x._data.shape[0]
    perm_order = np.empty([permutations, mat_n], dtype=np.int)
    for row in range(permutations):
        perm_order[row, :] = np.random.permutation(mat_n)

    permuted_stats = np.empty([permutations])
    mantel_perm_pearsonr_cy(x, perm_order, xmean, normxm,
                            ynorm, permuted_stats)

    return [orig_stat, permuted_stats]
```

The new code is harder to read, mostly due to the inlining of permutation and flattening logic, but it is also drastically faster. As can be seen from Table 2, the speedup ranges from about 100x on the AMD EPYC CPU to over 200x on the Intel Xeon CPU. We are also now able to compute the Mantel test in reasonable time on matrices of up to 100k*100k in size, which were previously considered intractable.

## 4.3 Validation costs

Before users can call either `pcoa` or `mantel`, they must create an appropriate object to hold the data buffer; in scikit-bio library this is implemented as a `DistanceMatrix` class. As part of the object creation process it verifies that the buffer is symmetric and hollow.

The used implementation is summarized in Algorithm 6. Once again, the algorithm used is very simple and elegant but not memory efficient. The matrix buffer is read several times from main memory and intermediate buffers are also created in the process.

---

Algorithm 6: Original distance matrix validation implementation

---

```python
import numpy as np

def is_symmetric_and_hollow(mat):
    # mat must be of type np.ndarray
    # and be a 2D floating point matrix
    not_sym = (mat.T != mat).any()
    not_hollow = (np.trace(mat) != 0)

    return [not no_sym, not not_hollow]
```

An optimized implementation uses both a tiled approach and fuses the two operations together. The cache-optimized version implemented in Cython is presented in Algorithm 7. The code is a little longer, but not significantly more complex.

---

Algorithm 7: Cache-optimized distance matrix validation implementation

---

```python
import numpy as np
from cython.parallel import prange

def is_symmetric_and_hollow_cy(float[:, :] mat):
    in_n = mat.shape[0]
    is_sym = True
    is_hollow = True

    # use a tiled approach to maximize memory locality
    for trow in prange(0, in_n, 16, nogil=True):
        trow_max = min(trow+16, in_n)
        for tcol in range(0, in_n, 16):
            tcol_max = min(tcol+16, in_n)
            for row in range(trow, trow_max, 1):
                for col in range(tcol, tcol_max, 1):
                    is_sym &= (mat[row,col]==mat[col,row])

            if (trow==tcol): # diagonal block
                for col in range(tcol, tcol_max, 1):
                    is_hollow &= mat[col,col]==0)

    return [(is_sym==True), (is_hollow==True)]
```

As expected, the new algorithm is significantly faster than the original. As can be seen from Table 3, the speedup ranges from about 5x on the AMD EPYC CPU to over 20x on the Intel Xeon.

Finally, having realized just how expensive validation really is, even after optimization, we added additional logic in the `DistanceMatrix` class to avoid unnecessary validations. For example, when making a copy of the object, one should be able to assume that the source was already validated and avoid re-validating it. This directly impacted `pcoa` performance, as it uses one object copy as part of its internal logic.

**Table 3: Runtimes for matrix validation, in seconds**

| CPU, number of used cores, matrix size | Original | Latest |
|---|---|---|
| AMD EPYC 7252 1 core - 25k x 25k | 1.8 | 2.6 |
| AMD EPYC 7252 8 cores - 25k x 25k | 1.8 | 0.4 |
| Intel Xeon Gold 6242 1 core - 25k x 25k | 4.4 | 3.2 |
| Intel Xeon Gold 6242 16 cores - 25k x 25k | 4.4 | 0.24 |
| AMD EPYC 7252 1 core - 70k x 70k | 18 | 21 |
| AMD EPYC 7252 8 cores - 70k x 70k | 18 | 3.2 |
| Intel Xeon Gold 6242 1 core - 100k x 100k | 219 | 78 |
| Intel Xeon Gold 6242 16 cores - 100k x 100k | 212 | 5.4 |

## 5 Summary and conclusion

This paper provides improved principal coordinate analysis and mantel test algorithms extensively used in the bioinformatics community. By taking into account the limited sizes of CPU caches and optimizing for memory locality, the new algorithms complete approximately 100x faster than their original implementations. The new algorithms maintain the same scikit-bio interface as the original ones, so no change in user code is needed to benefit from the improvements; the new code has been committed to the master branch of the library.

The effort of optimizing the above functions brought to light the limitations of over-reliance on standard libraries, especially in high-level languages like Python. The original code was implemented as a sequence of NumPy and SciPy matrix operations; while every single step was indeed high performance, the reliance of large intermediate buffers makes the algorithms completely memory-bound, and thus sub-optimal. When dealing with large buffers, focusing on memory locality and fusing as many operations as possible in the inner loop is the only option for fully exploiting the compute capabilities of modern CPUs. For Python-based programs, one way to accomplish this is to implement the double loop in Cython.

The scikit-bio library explicitly has the dual goals of providing easy to understand versions of algorithms for educational purposes, and optimized versions with the same API that run faster but are more difficult to understand. This work points to the utility of this strategy, but also to profile and further optimize the functionality that is widely used, as additional order-of-magnitude or more performance gains may be readily achievable in other areas. This would have a large impact in what analyses can be run on what kinds of systems, e.g. moving specific tasks off cluster or cloud environments and onto desktops or even phones, greatly broadening access and enabling new field applications that are hard to imagine today.

### ACKNOWLEDGMENTS

This work was partially funded by the US National Research Foundation (NSF) under grants DBI-2038509, OAC-1541349 and OAC-1826967.

# REFERENCES

[1]  Python Home Page, Online. https://www.python.org

[2]  NumPy Home Page. Online. https://numpy.org

[3]  Scikit-Bio Home Page. Online. http://scikit-bio.org

[4]  Evan Bolyen et. al. 2019. Reproducible, interactive, scalable and extensible microbiome data science using QIIME 2. Nature Biotechnology 37: 852–857. https://doi.org/10.1038/s41587-019-0209-9

[5]  Igor Sfiligoi, Daniel McDonald, and Rob Knight. 2020. Porting and Optimizing UniFrac for GPUs. PEARC '20: Practice and Experience in Advanced Research Computing. 500–504. https://doi.org/10.1145/3311790.3399614

[6]  Karim Esseghir. 1993. Improving Data Locality for Caches. Master's Thesis, Rice University. https://hdl.handle.net/1911/17049.

[7]  Stephanie Coleman and Kathryn S. McKinley. 1995. Tile Size Selection Using Cache Organization and Data Layout. Proc. SIGPLAN '95 Conf. Programming Language Design and Implementation, June 1995. https://doi.org/10.1145/223428.207162

[8]  Chris Mellor, Cascade Lake AP, Optane persistent memory and endurance. 2019. Blocks And Files. Online. https://blocksandfiles.com/2019/02/18/cascade-lake-ap-optane-persistent-memory-and-endurance/

[9]  Wikichip - Cascade Lakes, https://en.wikichip.org/wiki/intel/microarchitectures/cascade_lake

[10] Pierre Legendre and Louis Legendre. 1998. Numerical ecology, Developments in Environmental Modelling. Elsevier Science B.V.: Amsterdam.

[11] Nathan Halko, Per-Gunnar Martinsson, Yoel Shkolnisky, and Mark Tygert. 2011. SIAM J. Sci. Comput., 33(5), 2580–2594. (15 pages) https://doi.org/10.1137/100804139

[12] Cython Home Page. Online. https://cython.org

[13] "Pearson correlation coefficient", Wikipedia. Online. https://en.wikipedia.org/wiki/Pearson_correlation_coefficient

[14] Charles R. Harris et al. 2020. Array programming with NumPy. Nature 585, 357–362. https://doi.org/10.1038/s41586-020-2649-2

[15] Luke R. Thompson et al. 2017. A communal catalogue reveals Earth's multiscale microbial diversity. Nature 551, 457–463. https://doi.org/10.1038/nature24621

[16] Virtanen, P., Gommers, R., Oliphant, T.E. et al. SciPy 1.0: fundamental algorithms for scientific computing in Python. Nat Methods 17, 261–272 (2020). https://doi.org/10.1038/s41592-019-0686-2